\documentclass[aps,prb,showpacs,manuscript]{revtex4}
\usepackage{amsmath,amssymb}
\usepackage{graphicx}
\usepackage{dcolumn}
\usepackage{bm}
\usepackage{color}
\newcommand{\mbb}{\mathbb}
\newcommand{\mc}{\mathcal}

\newcommand{\tet}{\texttt}

\begin{document}

\title{Role played by strain on  plasmons, screening
and energy loss in graphene/substrate contacts}
  \author{ Dipendra Dahal$^1$, Godfrey Gumbs$^{1,2}$, Danhong Huang$^{3}$  }
 \affiliation{$^{1}$Department of Physics and Astronomy, Hunter College of the
City University of New York, 695 Park Avenue, New York, NY 10065, USA\\
$^{2}$Donostia International Physics Center (DIPC),
P de Manuel Lardizabal, 4, 20018 San Sebastian, Basque Country, Spain\\
 $^{3}$Air Force Research Laboratory, Space Vehicles Directorate, Kirtland Air Force Base, NM 87117, USA}

\date{\today}

\begin{abstract}
The combined effect due to mechanical strain, coupling to the plasmons in a
doped  conducting substrate, the plasmon-phonon coupling in conjunction
with the role played by encapsulation of a secondary two-dimensional (2D) layer
is investigated both theoretically and numerically. The calculations
are based on the random-phase approximation (RPA) for the surface response
function which yields the plasmon dispersion equation that is applicable in the
presence or absence of an applied uniaxial strain.  We present results showing
the dependence of the frequency of the charge density oscillations on the
strain modulus and direction of the wave vector in the Brillouin zone. The shielding
of a dilute distribution of charges as well as the rate of loss of energy   
for impinging charges is investigated for this hybrid layered structure.
\end{abstract}

\vskip 0.2in


\medskip
\par
\maketitle

\section{Introduction}
\label{sec1}

It is undoubtedly true that there has been a tremendous effort on the part of condensed
matter and materials scientists to increase their knowledge of the properties  of
low-dimensional structures. These include doped as well as undoped graphene,\cite{GR1,GR2,GR3} 
silicene,\cite{Si4,Si5} phosphorene,\cite{BP9,BP10}
germanene,\cite{Ge6,Ge7} antimonene,\cite{Sb11,Sb12} tinene,\cite{Ti8} 
bismuthene\cite{Bi13,Bi14,Bi15,Bi16,Bi17,Bi18} and most recently the  
two-dimensional pseudospin-1 $\alpha-T_3$ lattice\cite{Ejnicol}. Experimental studies of 
such structures may involve a wide range of techniques including
angle-resolved photoemission spectroscopy (ARPES){\cite {ARPES1,ARPES2,ARPES3,ARPES4}} and electron
energy loss spectroscopy (EELS){\cite{FMD2,Balassis,Zaremba, Gumbs}}.
Both of these methods rely on an analysis of the energy of an electron emitted from or
passing in the vicinity of the surface of the condensed matter under investigation.
Interest in these materials stems from their potential use in device applications including
transistors and state-of-the-art bismuth photonics as bismuth optical circuits
have emerged as a possible replacement technology for copper-based circuits in
communication and broadband networks.

\medskip
\par

\begin{figure}
\centering
\includegraphics[width=.45\textwidth]{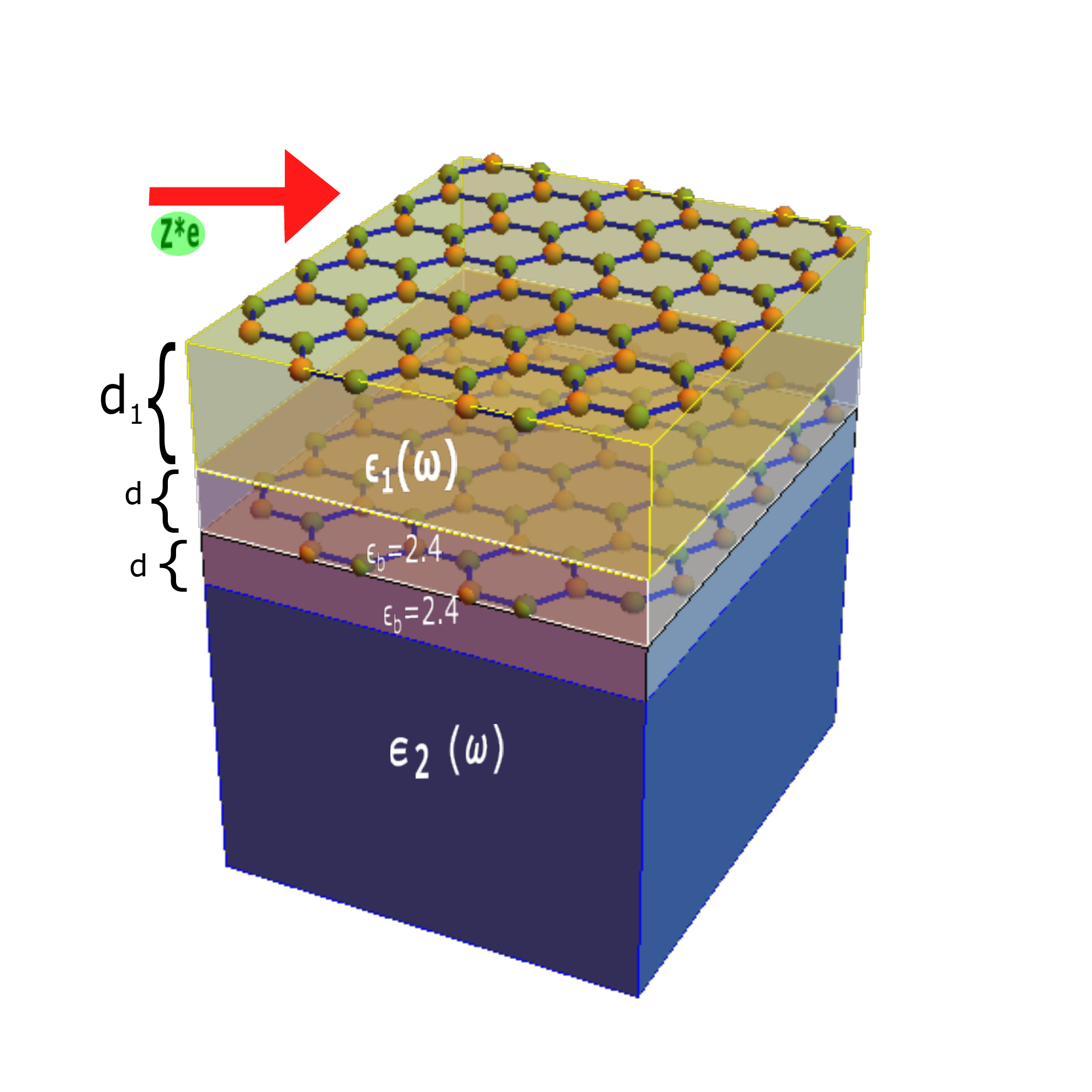}
\caption{(Color online)   Schematic illustration of a pair of
2D graphene layers nonlocally screened by two conducting materials
 with dielectric functions $\epsilon_1(\omega)$ and $\epsilon_2(\omega)$.
A background medium with dielectric constant $\epsilon_b=2.4$ lies between them.
A particle of charge $Z^\ast e$ moves parallel to the surface.}
\label{FIG:1}
\end{figure}

 We know that when an electromagnetic wave is incident on a material, especially on a conductor, the   quasiparticles can respond by oscillating at specific frequencies which could be sustained over considerable  distances and times if the frequency and wave number of the external perturbation are in resonance with the collective charge density oscillations.  Generally, the dispersion relation of these plasmon modes is determined by the geometric and electronic properties of the 2D layer as well as the nature of the conducting substrate  with which it is Coulomb coupled.   In the case of free-standing graphene, the frequency of the plasmon behaves as $\sqrt{q_\parallel}$ in the long wavelength limit.\cite{Wunsch}  However, the plasmon dispersion relation could be modified when the two-dimensional (2D) graphene sheet is subjected to strain and also if it is coupled to the charge density oscillations and plasmons in neighboring media as illustrated in Fig.\ \ref{FIG:1}.

\medskip
\par

Theoretical results for EELS have been presented for free-standing graphene in Ref.\ [\onlinecite{Balassis}] where the authors reported the contributions to the rate of loss of energy due to the single-particle and   plasmon excitations for particle motion parallel to the planar surface. The method of calculation was based on the formalism presented by previous authors \cite{TimTso,Com} who considered a 2D layer and cylindrical nanotube interacting with a beam of impinging charged particles.  However, in recent work, Woessner, et al. \cite{Vignale,Vignale2} released experimental and theoretical results for plasmon excitations in a heterostructure of graphene which is encapsulated\cite{encap1,encap2,pssb} between two films of hexagonal boron-nitride using a method that exploits near-field-microscopy.
The collective mode spectrum revealed
in the experimental data of Refs.\ [\onlinecite{Vignale,Vignale2}] is far more complex
than that in Ref.\ [\onlinecite{Balassis}] for free-standing graphene. Consequently,
we direct our attention to the heterostructure in Fig.\ \ref{FIG:1} which involves
atomically flat materials.  Our formalism includes contributions from plasmon-phonon 
coupling involving transverse and longitudinal optical phonons from the surrounding
conducting media.  Additionally, although there have been several papers dealing with   the effect due to mechanical strain  on the  plasmon  dispersion for free standing graphene by  Pelligrino et.al \cite{FMD1}, so far no consideration has been given to  the influence of strain on the fast-particle energy loss spectrum or the plasmon mode dispersion for structure coomposing of a 2D layer and conducting  substrate  for which  longitudinal and transverse phonon modes from the conducting substrate are taken into consideration.
\medskip
\par

When a graphene layer is subjected to mechanical strain, the regular crystal structure is deformed which leads to a modification of its energy band structure,\cite{FMD2,EB1,EB2,EB3,EB4,EB5,Jose} electrical and thermal conductivity,\cite{FMD2,cond1} as well as other transport properties.\cite{Trans1,Trans2}  Meanwhile, its polarizability is altered, thereby leading to qualitative changes in the plasmon mode dispersion relation. Making use of the polarization function derived in Refs.\ [\onlinecite{FMD1, FMD2,FMD3,FMD4}] for strained graphene,  we have investigated the plasmon mode dispersion for a structure  shown schematically in Fig.\ref{FIG:1}. In addition, we analyzed the effect due to  strain{\cite{Lin}} on the plasmon mode dispersion relation for previously studied structures\cite{Boperson, Horing} which are special cases of the illustrated hybrid heterostructure.  We have obtained analytical and numerical results showing the effect due to strain  and phonon vibrations in the substrate on the plasmon excitation spectrum in the long wavelength limit by varying  several  parameters including the angle giving the direction of the applied strain, the strain modulus, the separation  between the graphene layers, the dielectric constant for the background material and the wave vector.  This  information  will be  useful in  designing applications involving nanoelectronic and optoelectronic devices.

\medskip
\par

A critical ingredient which is needed for conducting our investigation outlined above is the surface response function.  This is achieved by using a transfer matrix method, as outlined in Ref.\
[\onlinecite {pssb}], involving the electrostatic potential, electric field and the induced charge density at the  interfaces of the structure shown in Fig.\  \ref{FIG:1}.  This procedure allows us to incorporate the effect due to energy band gap, mechanical strain,  as well as plasmon-phonon coupling,  all of which have not been investigated simultaneously so far in our hybrid structure. Additionally, we could exploit the calculated surface response function to determine the plasmonic and single-particle excitation contributions to the rate of loss of energy for a beam of charged particles moving in the vicinity of the heterostructure.

\medskip
\par

We have organized the rest of our paper as follows: In Sec. \ref{sec2}, we present the method for calculating the power loss of a charged particle and the introduction of the surface response function through the induced potential just outside the structure.  An explicit expression for the surface response function is obtained in  Sec. \ref{sec3} by ensuring the continuity of the electrostatic potential and accounting for the change in electric field due to the induced charge density on the 2D planes where conducting carriers are located.  In Sec.\  \ref{sec4}, closed-form analytic expressions are obtained in the
long wavelength limit for some specific geometrical arrangements arising from Fig.\ \ref{FIG:1}.  
Detailed analytical results for the plasmon dispersion relation are presented in Sec.\  
\ref{insert} for a pair of dissimilar 2D layers, with one acting as an overlayer for 
a dielectric in which the other is embedded.   This  arrangement is relevant to a 
recent experimental study of low frequency plasmons in a graphene-Cu(111) contact.
 Detailed numerical results for arbitrary wavelength are presented in Sec. \ref{sec5} showing  the combined  effect due to strain and plasmon-phonon coupling from the surrounding medium. 
Comparison of the energy loss from plasmons and single-particle excitations in strained and unstrained graphene is also presented. The versatility of the surface response function is further  demonstrated by calculating the screened potential of an impurity located near the surface of our hybrid structure. 
We concluded our paper with a brief summary of our accomplishments in Sec. \ref{sec6}. 

\medskip
\par

\medskip
\par

\section{Energy Loss in terms of the surface response function  }
\label{sec2}

We introduce our notation with a brief review. 
Let us assume that the medium occupies the half-space $z<0$.  Consider a point 
charge  $Z^\ast e$ moving along a prescribed path ${\bf r}(t)$ outside the medium.  
The external potential ${\phi_{ext}}({\bf r},t)$ satisfies Poisson's equation
$\nabla^2 \phi_{ext}({\bf r},t)=-  (Z^\ast e/\epsilon_0)\delta\left( {\bf r}-{\bf r}(t) 
\right) $  which has solution

\begin{equation}
\phi_{ext}({\bf r},t)=  \int \frac{d^2{\bf q}_\parallel}{(2\pi)^2}\int_{-\infty}^\infty d\omega\
\tilde{\phi}_{ext}\left({\bf q_\parallel},\omega\right)  e^{i({\bf q}_\parallel\cdot {\bf r}_\parallel  -\omega t)}
e^{q_\parallel z}\ ,
\end{equation}
where  $\tilde{\phi}_{ext}\left({\bf  q_\parallel},\omega\right)=- Z^\ast e /(4\pi\epsilon_0 q_\parallel)
{\cal F}\left({\bf q_\parallel},\omega  \right)$ 
  with the form factor defined as

\begin{equation}
{\cal F}\left({\bf q_\parallel},\omega  \right)\equiv  \int_0^\infty   dt\  e^{-q_\parallel z(t)}
e^{i(\omega t-{\bf q}_\parallel\cdot {\bf r}_\parallel(t) )} \ .
\label{Factor}
\end{equation}
In this notation, ${\bf q}_\parallel$ is a two-dimensional wave vector in the $xy$-plane parallel
to the surface which is situated at $z=0$. Also, it is understood that the frequency has a small imaginary part, i.e.,
$\omega\to \omega+i0^+$.

\medskip
\par

The external potential will give rise to an induced potential which, outside the
structure,  can be written as

\begin{equation}
\phi_{ind}({\bf r},t)=-\int \frac{d^2{\bf q}_\parallel}{(2\pi)^2}\int_{-\infty}^\infty d\omega\
\tilde{\phi}_{ext}\left( q_\parallel,\omega\right)  e^{i({\bf q}_\parallel\cdot {\bf r}_\parallel  -\omega t)}
g({\bf q_\parallel},\omega)e^{-q_\parallel z} \ .
\end{equation}
This equation defines the surface response function $g({\bf q_\parallel},\omega)$.  It has been implicitly assumed that
the external potential $\phi_{ext}$ is so weak that the medium responds linearly to it.  The function
$g({\bf q_\parallel},\omega)$ is itself related to the density-density response function
$\chi$ by

\begin{eqnarray}
g({\bf q_\parallel},\omega)&=& \frac{2\pi}{q_\parallel} \int_{-\infty}^\infty dz \int_{-\infty}^\infty dz^\prime
e^{q_\parallel z} e^{q_\parallel z^\prime} \
\chi(z,z^\prime ; {\bf q}_\parallel,\omega)
\nonumber\\
&=& - \int_{-\infty}^\infty dz \     e^{q_\parallel z} \rho_{ind}(z;{\bf q_\parallel},\omega)
\end{eqnarray}
which defines the induced surface charge density $\rho_{ind}(z;{\bf q_\parallel},\omega)$.

\medskip
\par

The quantity  Im $g({\bf q_\parallel},\omega)$ can be identified with the power absorption in the
structure  due to electron excitation induced by the external potential.  
The total potential in the vicinity of the surface ($z\approx 0$),  is given by

\begin{equation}
\phi ({\bf r},t) =     \int \frac{d^2{\bf q}_\parallel}{(2\pi)^2}\int_{-\infty}^\infty d\omega\
\left( e^{q_\parallel z}-g({\bf q_\parallel},\omega) e^{-q_\parallel z} \right)
 e^{i({\bf q}_\parallel\cdot {\bf r}_\parallel  -\omega t)}
\tilde{\phi}_{ext}\left( q_\parallel,\omega\right) 
 \label{eq:8}
\end{equation}
which takes account of nonlocal screening of the external potential.

\medskip
\par

Now, let us express the induced potential as

\begin{equation}
\phi_{ind} ({\bf r},t)=\frac{Z^\ast e}{4\pi\epsilon_0}  \int \frac{d^2{\bf q}_\parallel}{(2\pi)^2}
 \ \frac{1}{q_\parallel}\int_{-\infty}^\infty d\omega\
{\cal F}\left({\bf q_\parallel},\omega  \right)e^{i({\bf q_\parallel} \cdot {\bf r_\parallel}-\omega t)}
g\left({\bf q_\parallel},\omega  \right) e^{-{ q_\parallel} z}\ .
\end{equation}
Then, the  instantaneous force is

\begin{eqnarray}
{\bf F}_{ind}&=& e \nabla \left.\phi_{ind} ({\bf r},t)\right|_{{\bf r}={\bf r}(t)}
\nonumber\\
&=&   \frac{Z^\ast e^2}{4\pi\epsilon_0} \int \frac{d^2{\bf q}_\parallel}{(2\pi)^2} \
\frac{1}{q_\parallel}\int_{-\infty}^\infty d\omega\
{\cal F} \left(q_\parallel,\omega  \right)e^{i({\bf q}_\parallel \cdot {\bf r}_\parallel-\omega t)}
g\left(q_\parallel,\omega  \right) e^{-q_\parallel z}
\left.\left(i{\bf q}_\parallel -q_\parallel \hat{z}  \right) \right|_{{\bf r}={\bf r}(t)}\ .
\label{FINDU}
\end{eqnarray}
Assuming that the charge moves parallel to the surface with velocity ${\bf v}$ at a height $z_0$
so that its trajectory is described by ${\bf r}_\parallel (t)= {\bf v} t$  and $z(t)=z_0$.  Then,
in this case, the form factor in Eq.\ (\ref{Factor})  becomes
${\cal F}\left(q_\parallel,\omega  \right)=   i\ e^{-q_\parallel z_0}/  
( \omega-{\bf q}_\parallel\cdot {\bf v} ) $.  Making use of this result in Eq.\ (\ref{FINDU}),  a straightforward 
calculation yields  the rate of loss of energy of the charged particle to 
the medium of plasma as

\begin{equation}
\frac{dW}{dt}=
{\bf F}_{ind}\cdot {\bf v}=  \frac{Z^\ast e^2}{4\pi\epsilon_0} \int \frac{d^2{\bf q}_\parallel}{2\pi} \
\frac{{\bf q}_\parallel \cdot {\bf v}}{q_\parallel}
 e^{-2q_\parallel z_0} \ \mbox{Im}\ \left\{ g\left({\bf q_\parallel},\omega ={\bf q}_\parallel  \cdot  {\bf v}\right) \right\} \ .
\label{FINAL}
\end{equation}
We can use the result in Eq.\ (\ref{FINAL}) to determine the contributions to $dW/dt$ from the plasmon
excitations as well as the single-particle excitations for the hybrid structure shown schematically in Fig.\
\ref{FIG:1}. However, what is needed to proceed further with our calculation is an explicit formula for
$g\left({\bf q}_\parallel,\omega\right)$.  This is achieved by making sure that the potential just outside the surface at $z=0$ in Eq.\ (\ref{eq:8})   is continuous with that inside the material and the latter is continuous throughout  the $z<0$ region.

\section{Surface Response Function for a hybrid structure }
\label{sec3}

The structure shown schematically in Fig.  \ref{FIG:1} consists of a graphene layer on top of a conductor with dielectric function $\epsilon_1(\omega)$ and thickness $d_1$. This in turn lies on a dielectric with background constant $\epsilon_b$ and thickness $2d$ where another 2D layer is embedded in the middle. This whole structure is placed on a conducting substrate whose dielectric function is $\epsilon_2(\omega)$.    We write the potential in each region with a dielectric constant displayed in Fig.\ \ref{FIG:1}
as 

\begin{equation}
\phi_i ({\bf r},t) =   \int \frac{d^2{\bf q}_\parallel}{(2\pi)^2}\int_{-\infty}^\infty d\omega\
\left( a_i\ e^{-q_\parallel z}  +b_i e^{{q_\parallel} z} \right)
 e^{i({\bf q}_\parallel\cdot {\bf r}_\parallel  -\omega t)}
\tilde{\phi}_{ext}\left( q_\parallel,\omega\right)\ ,
\label{eq:8add}
\end{equation}
where $a_i,b_i$ are determined using the electrostatic conditions at the boundaries 
separating the regions.\cite{pssb}  After a straightforward calculation, 
we obtain the coefficients for the potential in the region $-d_1\leq z \leq0$ as

\begin{equation}
a_1=  -\ \frac{N_{11}({\bf q}_{\parallel},\omega)}{D_{11}({\bf q}_{\parallel},\omega)}  \ ,   
\  \  \ \  b_1=  \ \frac{N_{12}({\bf q}_{\parallel},\omega)}{D_{11}({\bf q}_{\parallel},\omega)}   \  ,
\end{equation}
where

\begin{equation}
N_{11}({\bf q}_{\parallel},\omega)=2q_{\parallel}\epsilon_0
\bigg\{e^{6d q_{\parallel}+4d_1q_{\parallel}} N_{a1}+2e^{4(d+d_1)
q_{\parallel}}N_{a2}-e^{2(d+2d_1)q_{\parallel}}N_{a3}\bigg\}
\end{equation}
and

\begin{eqnarray}
N_{a1}&=&(\text{$\epsilon_1(\omega)$}-\text{$\epsilon_b$}) (\text{$\epsilon_2(\omega)$}+\text{$\epsilon_b$}) (2 q_{||} \text{$\epsilon_0$} \text{$\epsilon_b$}-\text{$\chi_2$})\ ,\nonumber\\
N_{a2}&=&\text{$\chi_2$} \left(\text{$\epsilon_1(\omega)$} \text{$\epsilon_2(\omega)$}+\text{$\epsilon_b$}^2\right)\ ,\nonumber\\
N_{a3}&=&(\text{$\epsilon_1(\omega)$}+\text{$\epsilon_b$}) (\text{$\epsilon_2(\omega)$}-\text{$\epsilon_b$}) (2 q_{||} \text{$\epsilon_0$} \text{$\epsilon_b$}+\text{$\chi_2$})\ .
\end{eqnarray}
Also,

\begin{equation}
N_{12}({\bf q}_{\parallel},\omega)=-\ 2e^{2d_1q_{\parallel}}q_{\parallel}
\epsilon_0 \bigg\{e^{6dq_{\parallel}+4d_1q_{\parallel}}N_{b1}+
2e^{4(d+d_1)q_{\parallel}}N_{b2}-  e^{2(d+2d_1)q_{\parallel}}N_{b3}\bigg\}
\end{equation}
with

\begin{eqnarray}
N_{b1}({\bf q}_{\parallel},\omega)&=&(\epsilon_1(\omega)+\epsilon_b)( \epsilon_2(\omega)+
\epsilon_b) (2 q_{\parallel} \epsilon_0 \epsilon_b-
\chi_2({\bf q_\parallel},\omega)\ ,
\nonumber\\
N_{b2}({\bf q}_{\parallel},\omega)&=&\chi_2({\bf q_\parallel},\omega) 
\left(\epsilon_1(\omega) \epsilon_2(\omega)-\epsilon_b^2\right)\ ,
\nonumber\\
N_{b3}({\bf q}_{\parallel},\omega)&=&(\epsilon_1(\omega)-\epsilon_b) 
(\epsilon_2(\omega)-\epsilon_b )(2 q_{\parallel} \epsilon_0 
\epsilon_b+\chi_2({\bf q_\parallel},\omega)
\end{eqnarray} 
and

\begin{eqnarray}    
D_{11}({\bf q}_{\parallel},\omega)=&&e^{6(d+d_1)q_{\parallel}}
D_{d1}+e^{6 d q_{\parallel} + 4 d_1 q_{\parallel}}D_{d2}
    -2 e^{(4 d q_{\parallel} + 6 d_1 q_{\parallel})}D_{d3}
		\nonumber\\
&&+ 2 e^{(4 (d + d_1) q_{\parallel})}D_{d4}+e^{(2 (d + 3 d_1) q_{\parallel})}
D_{d5}-e^{(2 (d + 2 d_1) q_{\parallel})}D_{d6}
\end{eqnarray}
with

\begin{eqnarray}
D_{d1}({\bf q}_{\parallel},\omega)&=&(\text{$\epsilon_1(\omega)$}+\text{$\epsilon_b$}) (\text{$\epsilon_2(\omega)$}+\text{$\epsilon_b$}) \big\{q_{\parallel} \text{$\epsilon_0$} (\text{$\epsilon_1(\omega)$}+1)-\text{$\chi_1$}\big\} (2 q_{\parallel} \text{$\epsilon_0$} \text{$\epsilon_b$}-\text{$\chi_2$})\ ,\nonumber\\
D_{d2}({\bf q}_{\parallel},\omega)&=&(\text{$\epsilon_1(\omega)$}-\text{$\epsilon_b$}) (\text{$\epsilon_2(\omega)$}+\text{$\epsilon_b$}) \big\{q_{\parallel} \text{$\epsilon_0$} (\text{$\epsilon_1(\omega)$}-1)+\text{$\chi_1$}\big\} (2 q_{\parallel} \text{$\epsilon_0$} \text{$\epsilon_b$}-\text{$\chi_2$})\ ,\nonumber\\
D_{d3}({\bf q}_{\parallel},\omega)&=&\text{$\chi_2$} \left(\text{$\epsilon_1(\omega)$} \text{$\epsilon_2(\omega)$}-\text{$\epsilon_b$}^2\right) \big\{q_{\parallel} \text{$\epsilon_0$} (\text{$\epsilon_1(\omega)$}+1)-\text{$\chi_1$}\big\}\ ,\nonumber\\
D_{d4}({\bf q}_{\parallel},\omega)&=&\text{$\chi_2$} \left(\text{$\epsilon_1(\omega)$} \text{$\epsilon_2(\omega)$}+\text{$\epsilon_b$}^2\right) \big\{q_{\parallel} \text{$\epsilon_0$} (\text{$\epsilon_1(\omega)$}-1)+\text{$\chi_1$}\big\}\ ,\nonumber\\
D_{d5}({\bf q}_{\parallel},\omega)&=&(\text{$\epsilon_1(\omega)$}-\text{$\epsilon_b$}) (\text{$\epsilon_2(\omega)$}-\text{$\epsilon_b$}) \big\{q_{\parallel} \text{$\epsilon_0$} (\text{$\epsilon_1(\omega)$}+1)-\text{$\chi_1$}\big\} (2 q_{\parallel} \text{$\epsilon_0$} \text{$\epsilon_b$}+\text{$\chi_2$})\ ,\nonumber\\
D_{d6}({\bf q}_{\parallel},\omega)&=&(\text{$\epsilon_1(\omega)$}+\text{$\epsilon_b$}) (\text{$\epsilon_2(\omega)$}-\text{$\epsilon_b$}) \big\{q_{\parallel} \text{$\epsilon_0$} (\text{$\epsilon_1(\omega)$}-1)+\text{$\chi_1$}\big\} (2 q_{\parallel} \text{$\epsilon_0$} \text{$\epsilon_b$}+\text{$\chi_2$}) \ ,
\end{eqnarray}
where the (${\bf q}_\parallel ,\omega$)-dependence of the layer susceptibilities $\chi_1$ and $\chi_2$ 
has been suppressed for  convenience.  Additionally,  the surface response function is expressed as:   


 \begin{equation}
g({\bf q}_{\parallel},\omega)=\frac{{\cal N}({\bf q}_{\parallel},\omega)}{{\cal D}({\bf q}_{\parallel},\omega)} 
\label{gfn}
\end{equation}
with

\begin{eqnarray}
{\cal N}({\bf q}_{\parallel},\omega)=&&\left\{q_{\parallel}\epsilon_0  \left[ \epsilon_1(\omega)-1\right]-
\chi_1({\bf q}_{\parallel},\omega) \right\} \left[A_1+A_3+A_5 \right] +\left\{ q_{\parallel}\epsilon_0
\left[ \epsilon_1(\omega)+1 \right]
+\chi_1({\bf q}_{\parallel},\omega) \right\}\nonumber\\ &&\times[-A_2-A_4+A_6]\ ,
\end{eqnarray}

\begin{eqnarray}
{\cal D}({\bf q}_{\parallel},\omega)=&&\left\{q_{\parallel}\epsilon_0  \left[ \epsilon_1(\omega)+1\right]-
\chi_1({\bf q}_{\parallel},\omega) \right\} \left[A_1+A_3+A_5 \right] +\left\{ q_{\parallel}\epsilon_0
\left[ \epsilon_1(\omega) -1 \right]
+\chi_1({\bf q}_{\parallel},\omega) \right\}\nonumber\\ &&\times[-A_2-A_4+A_6]
\label{Denm}
\end{eqnarray}
we have

\begin{eqnarray}
A_1({\bf q}_{\parallel},\omega)&=&  e^{8(d_1+d)q_{\parallel}}
\left[ \epsilon_1(\omega)+\epsilon_b\right]
\left[\epsilon_2(\omega)+ \epsilon_b\right]  \left[ 2 q_{\parallel}\epsilon_0\epsilon_b-
\chi_2({\bf q}_{\parallel},\omega)\right] \ ,
\nonumber\\
A_2({\bf q}_{\parallel},\omega)&=& e^{2(3d_1+4d)q_{\parallel}}\left[ \epsilon_1(\omega)-
\epsilon_b\right]
\left[\epsilon_2(\omega)+ \epsilon_b\right]  \left[ 2 q_{\parallel}\epsilon_0\epsilon_b-
\chi_2({\bf q}_{\parallel},\omega)\right] \ ,
\nonumber\\
A_3({\bf q}_{\parallel},\omega)&=&2  e^{2(4d_1+3d)q_{\parallel}}\left[  \epsilon_1(\omega)
\epsilon_2(\omega)-\epsilon_b^2
\right] \chi_2({\bf q}_{\parallel},\omega) \ ,
\nonumber\\
A_4({\bf q}_{\parallel},\omega)&=&2 e^{6(d_1+d)q_{\parallel}}\left[\epsilon_1(\omega)\epsilon_2(\omega)+\epsilon_b^2
\right] \chi_2({\bf q}_{\parallel},\omega) \ ,
\nonumber\\
A_5({\bf q}_{\parallel},\omega)&=&e^{4(2d_1+d)q_{\parallel}}\left[\epsilon_1(\omega)-
\epsilon_b\right]\left[-\epsilon_2(\omega)+\epsilon_b
\right]\left[2q_{\parallel}\epsilon_0 \epsilon_b+\chi_2({\bf q}_{\parallel},\omega)\right] \ ,
\nonumber\\
A_6({\bf q}_{\parallel},\omega)&=&e^{2(3d_1+2d)q_{\parallel}} \left[\epsilon_2(\omega)-\epsilon_b\right]
\left[\epsilon_1(\omega)+\epsilon_b\right] \left[2q_{\parallel}\epsilon_0
\epsilon_b+\chi_2({\bf q}_{\parallel},\omega) \right] \ .
\end{eqnarray}
In this notation, $\epsilon_0$ is the permittivity  of free space,
for the upper  2D layer, we write for convenience  $ \chi_1({\bf q}_{\parallel},\omega)=
e^2 \Pi_1({\bf q}_{\parallel},\omega)$
and, similarly, for the lower layer,  $ \chi_2({\bf q}_{\parallel},\omega)=
e^2 \Pi_2({\bf q}_{\parallel},\omega)$. Here,   $e$ is the electron charge and,
for convenience, we have introduced the polarization functions
 $\Pi_{1}({\bf q}_{\parallel},\omega),\   \Pi_{2}({\bf q}_{\parallel},\omega)$.
As a matter of fact, we have

\begin{equation}
\Pi({\bf q},\omega) = 
\int  \frac{d\omega^\prime d{\bf k}}{i(2\pi)^3}\mbox{Tr}
\left[ G^0({\bf k},\omega^\prime)G^0({\bf k}+{\bf q},\omega+\omega^\prime) \right]\  ,
\label{pi0}
\end{equation}
where $G^0({\bf k},\omega)$ is a single-particle Green's function
which is a $2\times 2$ matrix due to the underlying $A$ and $B$ sublattices.

\medskip
\par
The low-energy model Hamiltonian for unstrained graphene is well known and given by
${\bf H}^{(0)}=\hbar v_F {\bf \sigma} \cdot{\bf q}$
where $v_F$ is the Fermi velocity, ${\bf \sigma}=\{\sigma_x,\sigma_y\}$ in terms of
 Pauli matrices. When strain is applied, the low-energy Hamiltonian can be written as

\begin{equation}
{\bf H}=\hbar v_F {\bf \sigma} \cdot {\bf q}^\prime
\end{equation}
with ${\bf q}^\prime =\tensor{R}(\theta)\tensor{S}(\zeta)\tensor{R}(-\theta){\bf q}
=(\tensor{I}-2\kappa\tensor{\zeta}){\bf q}$,

\begin{equation}
\tensor{\zeta}=\zeta \begin{pmatrix}\cos^2\theta-\nu \sin^2\theta& (1+\nu)\cos\theta \sin\theta\\
(1+\nu)\cos\theta \sin\theta& \sin^2\theta-\nu \cos^2\theta
\end{pmatrix},
\end{equation}
and  $\tensor{I}$ is the unit $2 \times 2$ matrix, $\tensor{S}(\zeta)
=\mbox{diag}(c_{\parallel},c_{\perp})$,
$\kappa=\frac{a}{2t}|\frac{\partial t}{\partial a}|-\frac{1}{2}\approx 1.1$, the carbon carbon
bond length is $a=1.42 A^\circ$,  $R(\theta)$ as the rotation matrix in the direction of the applied strain and $\theta$ as the angle of the applied strain with respect to the $x$-axis,  the known value for Poisson's ratio for graphite is $\nu =0.165$ and for monolayer graphene it is $\nu$ as $0.14$. The difference between the  two values for the Poisson ratio is negligible compared with other parameters in our calculation. However, we chose the former value because the graphene sheet is part of a multi-layer structure.
We  have

\begin{equation}
\begin{pmatrix}{\bf q_x'}\\ {\bf q_y'}\end{pmatrix}=\begin{pmatrix}{\bf q_x}-2\kappa \zeta_{xx}{\bf q_x}-2\kappa \zeta_{xy}{\bf q_y} \\ {\bf q_y}-2\kappa \zeta_{yx}{\bf q_x}-2\kappa \zeta_{yy}{\bf q_y}\end{pmatrix}\ .
\end{equation}

\medskip
\par

Defining   the eigenvalues and eigenvectors in the pseudospin space of the Hamiltonian without and with applied strain, as $H^{(0)}|q',\pm>^{(0)}=E^{(0)}_{\pm q'}|q',\pm>^{(0)}$ and $H|q,\pm>=E|q,\pm>$, respectively, with $\pm $ as a pseudospin index, it follows that both $E_\pm q$ and $|q,\pm>$ under applied strain are mapped onto $E^{(0)}_{\pm q'}$ and $|{\bf q'},\pm>$. The polarization function of strained graphene would then be mapped onto the polarization function of unstrained graphene by \cite{Wunsch}

\begin{equation}
\Pi({\bf q},\omega)=\frac{1}{\mbox{Det}\ S(\zeta)}   \Pi^{(0)}({\bf q'},\omega)  \  ,
\end{equation}
where $\Pi^{(0)}({\bf q'},\omega)$ is the polarizability of unstrained monolayer graphene.
For small values of strain on graphene, the generalized polarization function {\cite {FMD1,FMD3}} may be obtained from a Taylor series expansion in $\zeta$ and expressed approximately as

\begin{eqnarray}
\Pi({\bf q}_\parallel,\omega) = &&\left[1+2\kappa(1-\nu)\zeta   \right] \Pi^{(0)}({\bf q}_\parallel,\omega)
-2\kappa   \frac{\partial  \Pi^{(0)}({\bf q}_\parallel,\omega)}{\partial  q_h}\zeta_{hk}q_k\nonumber\\
+&&2\kappa^2 [1+2\kappa \zeta(1-\nu)] \bigg[\frac{\partial^2}{\partial q_x^2}\Pi^0(q_x,q_y,\omega)\big(\zeta_{xx}q_x+\zeta_{xy}q_y\big)^2\nonumber\\
+&&2\frac{\partial^2}{\partial q_x \partial q_y}\Pi^0(q_x,q_y,\omega) (\zeta_{xx}q_x+\zeta_{xy}q_y)(\zeta_{yx}q_x+\zeta_{yy}q_y)\nonumber\\
+&& \frac{\partial^2}{\partial q_y^2}\Pi^0(q_x,q_y,\omega)(\zeta_{yx}q_x+\zeta_{yy}q_y)^2\bigg]\ .
\label{strain pi}
\end{eqnarray}
The subindex $h, k$ denotes  $x$ and $y$ and the summation convention is adopted here. With the aid of  the expression for the  polarization of unstrained monolayer graphene in Ref. [{\onlinecite {Wunsch}}] one could proceed to  calculate plasmon excitations in dimensionally mismatched Coulomb coupled 2D  systems using the obtained surface response function. However, before we do so, we will examine from a numerical point of view the effect of strain on the polarization function.

\medskip
\par

\begin{figure}
\centering
\includegraphics[width=.65\textwidth]{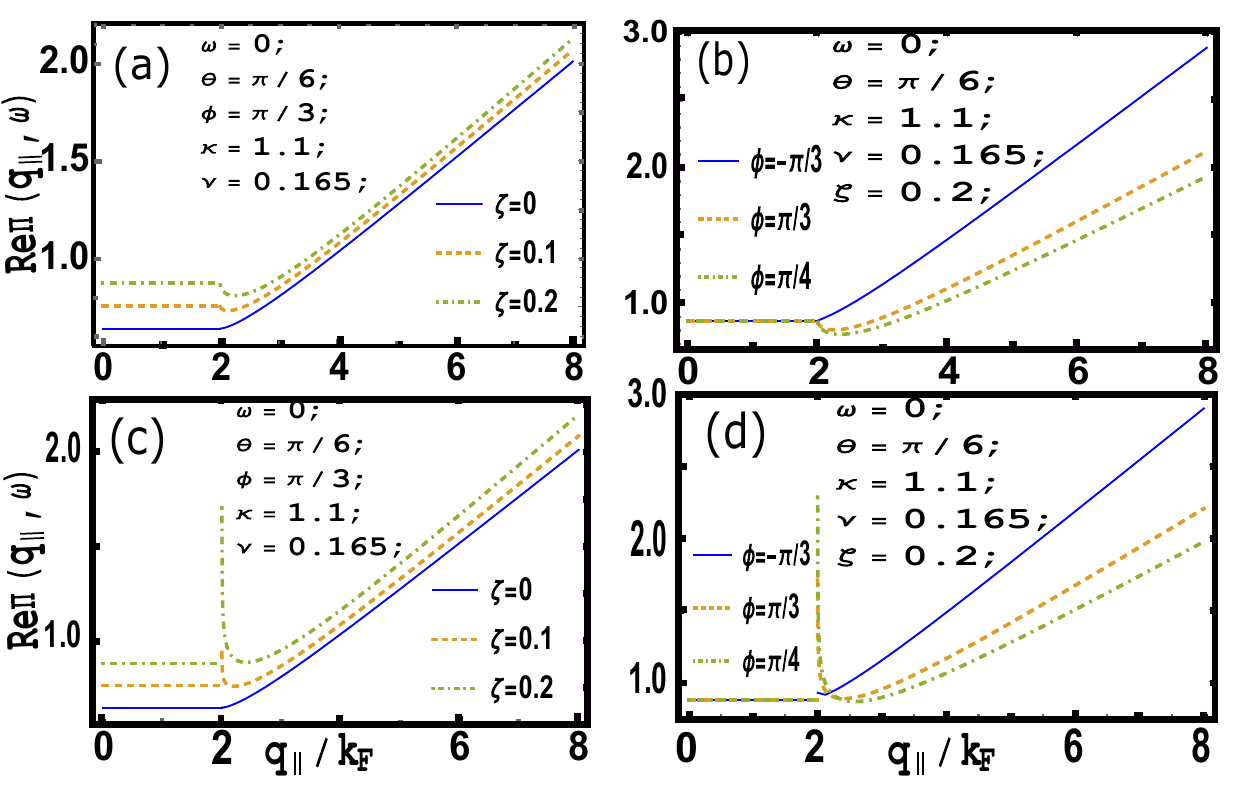}
\caption{
(Color online) Plots showing the real part of the polarization for (a) various values of strain and (b) various direction of wave vector when only the first order correction term in strain is included.  Panels (c) and (d) show the polarization for strained graphene for various strain and wave vector directions, respectively, when both first and second order corrections terms are included. The values for the  other parameters are shown in the figures.}
\label{POLZN}
\end{figure}

Making use of the expression for the  polarization function given in Eq. (\ref{strain pi}),  by including or neglecting the second-order correction term, we obtain the  behavior  of the real part of the static  polarization as shown in Fig.\ \ref{POLZN} . The upper panels of the figure show the polarizability  when only the first-order correction term is included and the bottom panels correspond to the polarization when both first and second-order terms contribute. Figure \ \ref{POLZN}(a) shows the polarization for three values of strain. For chosen  strain, the polarization remains constant in the range $0\leq q_{\parallel}<2k_F$. At $q_{\parallel}=2k_F$, we see a dip  due to the strain which monotonically increases afterwards. The magnitude of the polarization and the size of dip increases with increasing value of  strain.
 Figure\ \ref{POLZN}(b) in the top  right panel shows the variation of polarization due to change in wave vector direction. There, we see the polarization remaining constant in the range $0\leq q_{\parallel}<2k_F$ and has the same value for any direction of the wave vector. However, the value changes when the wave vector exceeds  twice the Fermi wave vector. The polarization value increases monotonically outside this range of wave vector. We could see similar behavior in the bottom panel figures when the second-order correction terms are considered. The main difference that we see there is the discontinuity at $q_\parallel=2k_F$ when strain is applied. This is due to the indeterminate nature of polarization at $q_{\parallel}=2k_F$.

\section{Dispersion relation for strained 2D layer-dielectric-conducting substrate
heterostructure}
\label{sec4}

We now turn our attention to a detailed study when a 2D layer is at a distance $d_1$
from a semi-infinite
conducting substrate with a dielectric function $\epsilon_2(\omega)$ with the space in between them filled with a medium of dielectric constant $\epsilon_b$. For this case, we replace
$\epsilon_1(\omega)$ in Eq. (\ref{gfn}) by $\epsilon_b$, set $\chi_2$ and $d$ equal to zero.
The resulting surface response function becomes

\begin{equation}
g_{Hybrid}(q_{\parallel},\omega)=\frac{N_{ Hybrid}(q_\parallel,\omega)}{D_{ Hybrid}(q_\parallel,\omega)} \  ,
\label{DDHC1}
\end{equation}
where

\begin{equation}
N_{ Hybrid}(q_\parallel,\omega)=(\epsilon_b-1)\{1+\frac{(\epsilon_2-\epsilon_b)}{(\epsilon_2+\epsilon_b)}\frac{(\epsilon_b+1)}{(\epsilon_b-1)}e^{-2q_{\parallel}d_1}\}-2\frac{\chi_1}{2q_{\parallel}\epsilon_0}\{1-\frac{\epsilon_2-\epsilon_b}{\epsilon_2+\epsilon_b}e^{-2q_{\parallel}d_1}\}\ ,
\end{equation}

\begin{equation}
D_{ Hybrid}(q_\parallel,\omega)=(\epsilon_b+1)\{1+\frac{(\epsilon_2-\epsilon_b)}{(\epsilon_2+\epsilon_b)}\frac{(\epsilon_b-1)}{(\epsilon_b+1)}e^{-2q_{\parallel}d_1}\}-2\frac{\chi_1}{2q_{\parallel}\epsilon_0}\{1-\frac{(\epsilon_2-\epsilon_b)}{(\epsilon_2+\epsilon_b)}e^{-2q_{\parallel}d_1}\}
\label{NHPRB}
\end{equation}
and we shall set $\chi_1=e^2\Pi({\bf q}_\parallel,\omega)$.

\medskip
\par

\begin{figure}
\centering
\includegraphics[width=0.75\textwidth]{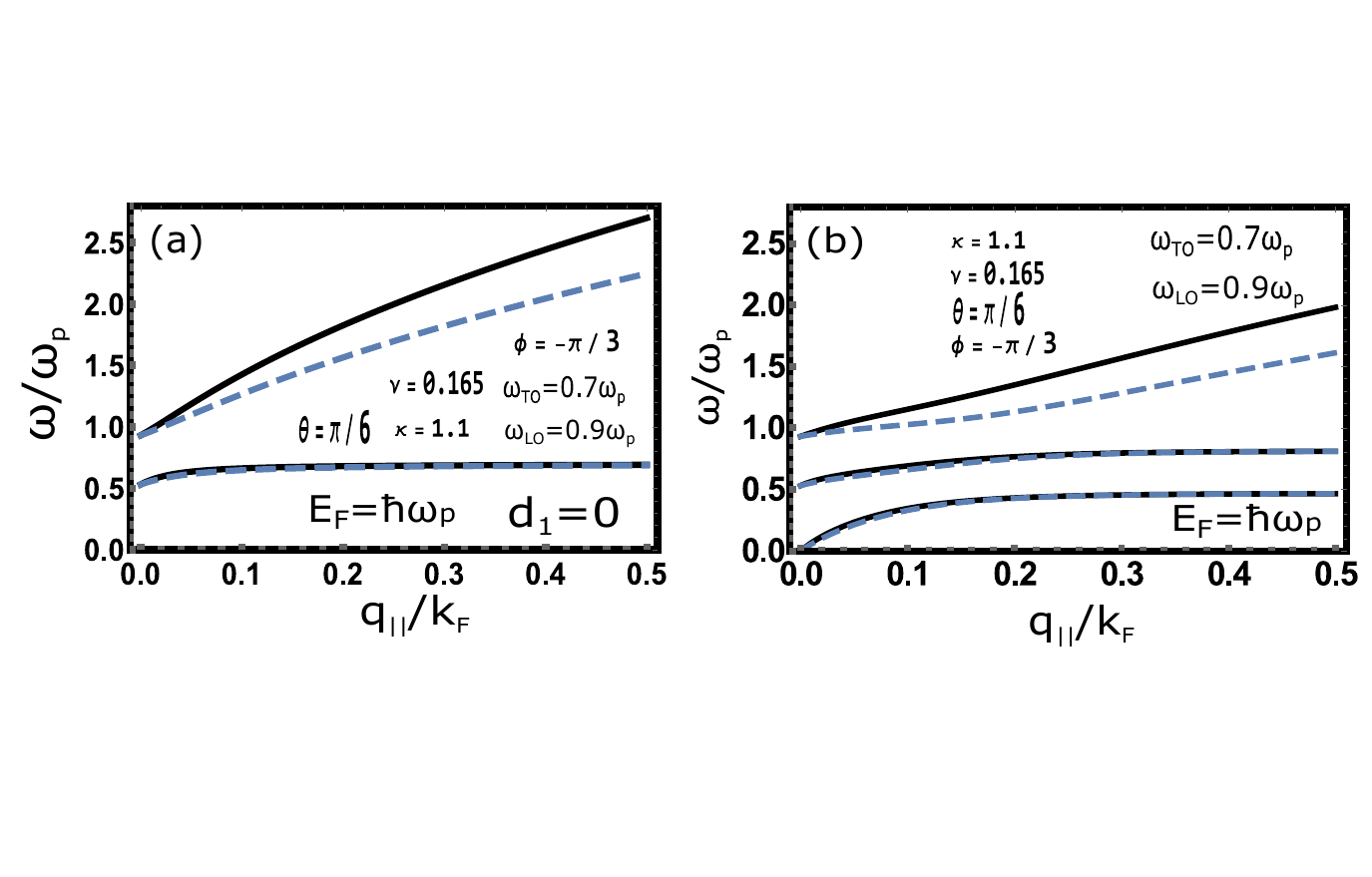}
\caption{(Color online)   Plasmon mode dispersion relation
 for gapless graphene in the long wavelength limit in the presence(solid curve) and absence(dashed curve) of strain.
These results demonstrate the effect due to plasmon-phonon interaction.
In (a), there is no separation between the 2D layer and the substrate. In
(b),  this separation is finite and chosen as $d_1=5.0k_F^{-1}$. }
\label{FIG: plasphoint}
\end{figure}

At long wavelengths, we have

\begin{equation}
\Pi(q_{\parallel},\omega) \approx\frac{2 E_F}{\pi \hbar^2}B(\theta,\phi) \frac{q_{\parallel}^2}{\omega^2}\ .
\end{equation}
\medskip
Making use of this approximation for the polarizability  in Eq.\ (\ref{NHPRB})  and then
setting the resulting equation equal to zero, we obtain the
dispersion equation for plasma excitations as

\begin{equation}
(\text{$\epsilon_b $}+1) \left\{e^{-2q_{\parallel}d_1}\frac{(\text{$\epsilon_b $}-1) (\text{$\epsilon_2 $}-\text{$\epsilon_b $})}{(\text{$\epsilon_b $}+1) (\text{$\epsilon_2 $}+\text{$\epsilon_b $})}+1\right\}-\frac{\text{K} \text{$q_{\parallel}$} B(\theta ,\phi ) \left\{\frac{(\text{$\epsilon_b $}-\text{$\epsilon_2 $}) e^{-2 \text{$d_1$} \text{$q_{\parallel}$}}}{\text{$\epsilon_2 $}+\text{$\epsilon_b $}}+1\right\}}{\text{$\omega$}^2}=0 \ ,
\label{Dzeroeqn}
\end{equation}
where $K= 2E_F e^2/(\pi \epsilon_0 \hbar^2)$ and
$B(\theta,\phi)=1-2 \kappa (1+\nu)\zeta cos2(\theta-\phi)$ with
$\phi$, indicating the direction of the wave vector.
What remains to be specified for solving Eq. \  (\ref{Dzeroeqn}) is the form for $\epsilon_2(\omega)$.  In accounting for coupling between the plasmons in the 2D layer with those with frequency $\omega_{p}$ in the conducting substrate as well as the longitudinal and transverse optical phonons with frequency  $\omega_{LO}$ and $\omega_{TO}$, respectively, in this case, we have

\begin{equation}
\epsilon_2(\omega)=1+\frac{\omega_{LO}^2-\omega_{TO}^2}{\omega_{TO}^2-\omega^2}-
\frac{\omega_p^2}{\omega^2} \  .
\label{plasmonphonon}
\end{equation}
The analytic solution of Eq.  (\ref{Dzeroeqn}) in conjunction with Eq.\ (\ref{plasmonphonon}) for $\omega$ is  unwieldy  and is not suitable for presentation.  Consequently, we present numerical results for the plasmon dispersion relations in Figs. \ref{FIG: plasphoint}(a) 
and \ref{FIG: plasphoint}(b) where we compare strained and unstrained graphene. In Fig. \ref{FIG: plasphoint}(a), there is no separation between the 2D layer and the surface ($d_1=0$), whereas in Fig. \ref{FIG: plasphoint}(b), there is a separation ($d_1=5.0k_F^{-1}$).   This difference leads to a semi-linear plasmon branch originating from the origin in Fig. \ref{FIG: plasphoint}(b).  In both panels, there is a plasmon branch close to  $0.5 \omega_p$ and another near  $1.0 \omega_p$ when $q_\parallel\to 0$.
These two plasmon branches are a direct consequence of the plasmon-phonon interaction. Finite separation of 2D layer and the conducting substrate generates new plasmon branch from the origin called Acoustic plasmon branch. Also, for all branches in strained graphene, the slope of the uppermost dispersion curve increases the most as the strain is increased whereas, in contrast,  the effect on the two other lower branches is small. This indicates how the plasmon frequency and its group velocity may be tuned for device applications.
			
\medskip
\par

The algebra involved in solving Eq.\ (\ref{Dzeroeqn})  is considerably simplified if we neglect the plasmon-phonon coupling and instead use  $\epsilon_2(\omega)=1- \omega_p^2/\omega^2$.  After
a straightforward calculation, we obtain

\begin{equation}
\omega_{\pm}(q_{\parallel},\theta,\phi)=\left\{\frac{A_1(q_{\parallel},\theta,\phi)
\pm\sqrt{A_1(q_{\parallel},\theta,\phi)^2-4 N_1(q,\theta,
\phi)}}{-2(-1+\epsilon_b)^2+2 e^{q_{\parallel}d_1}(1+\epsilon_b)^2}\right\}^{1/2} \ ,
\end{equation}
where

\begin{equation}
A_1(q_{\parallel},\theta,\phi)=\big\{B(\theta,\phi)Kq_{\parallel}
+\omega_p^2\big\}\big\{-1+\epsilon_b+e^{2q_{\parallel}d_1}(1+\epsilon_b)\big\}\ ,
\end{equation}

\begin{equation}
N_1(q_{\parallel},\theta,\phi)=B(\theta,\phi)Kq_{\parallel}\omega_p^2
(-1+e^{2q_{\parallel}d_1})\bigg\{-1+\epsilon_b+e^{2q_{\parallel}d_1}(1+\epsilon_b)\bigg\}\ .
\end{equation}

\medskip
\par

As a special case that is of interest to experimentalists, we consider  $SiO_2$ as the dielectric background which has dielectric constant, $\epsilon_b=3.8$.\cite{Peeters} The corresponding dispersion relations for this structure in the long wavelength limit are given by

\begin{equation}
\omega_1(q_{\parallel},\theta,\phi) \approx \frac{\omega_p}{\sqrt{2}}+\frac{1}{\sqrt{2}\omega_p}
\left\{\frac{K B(\theta, \phi)}{2}-\frac{84 d_1 \omega_p^2}{95}\right\}\  q_{\parallel} \  ,
\label{number1}
\end{equation}
and

\begin{equation}
\omega_2(q_\parallel,\theta,\phi) \approx  \left\{\frac{5}{19}B(\theta,\phi)d_1K 
\right\}^{1/2}q_\parallel \  .
\label{number2}
\end{equation}

\begin{figure}
\centering
\includegraphics[width=0.55\textwidth]{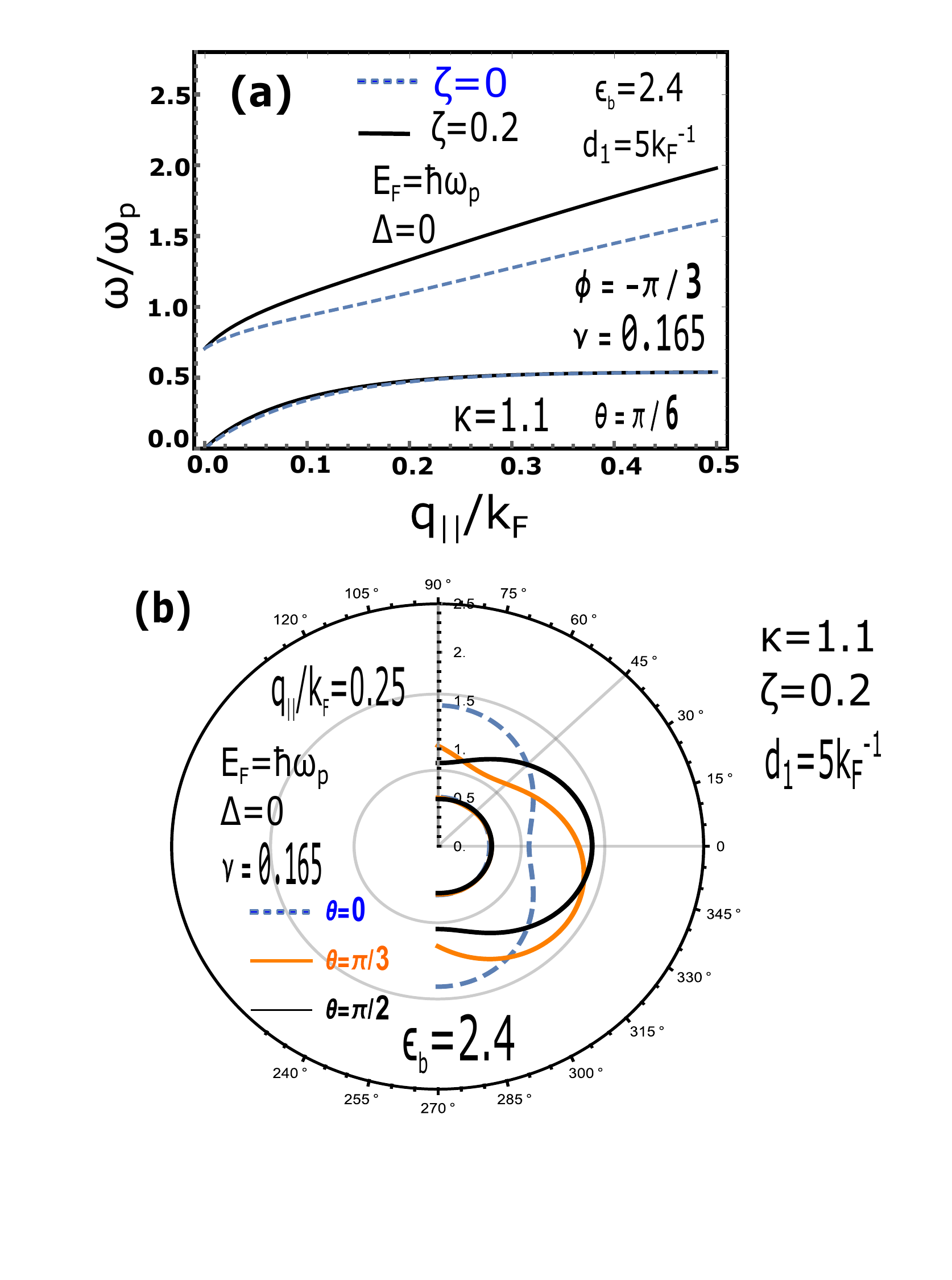}
\caption{(Color online)   Plasmon mode dispersion relation
 for gapless graphene in the  long wavelength approximation.  (a) The plasmon modes dispersion for strained and unstrained graphene and (b)  \  polar plots showing the variation of the plasmon modes for chosen wave vector. The electron-phonon coupling is neglected.}
\label{FIG:Hornopho}
\end{figure}

\medskip
\par

\begin{figure}
\centering
\includegraphics[width=.75\textwidth]{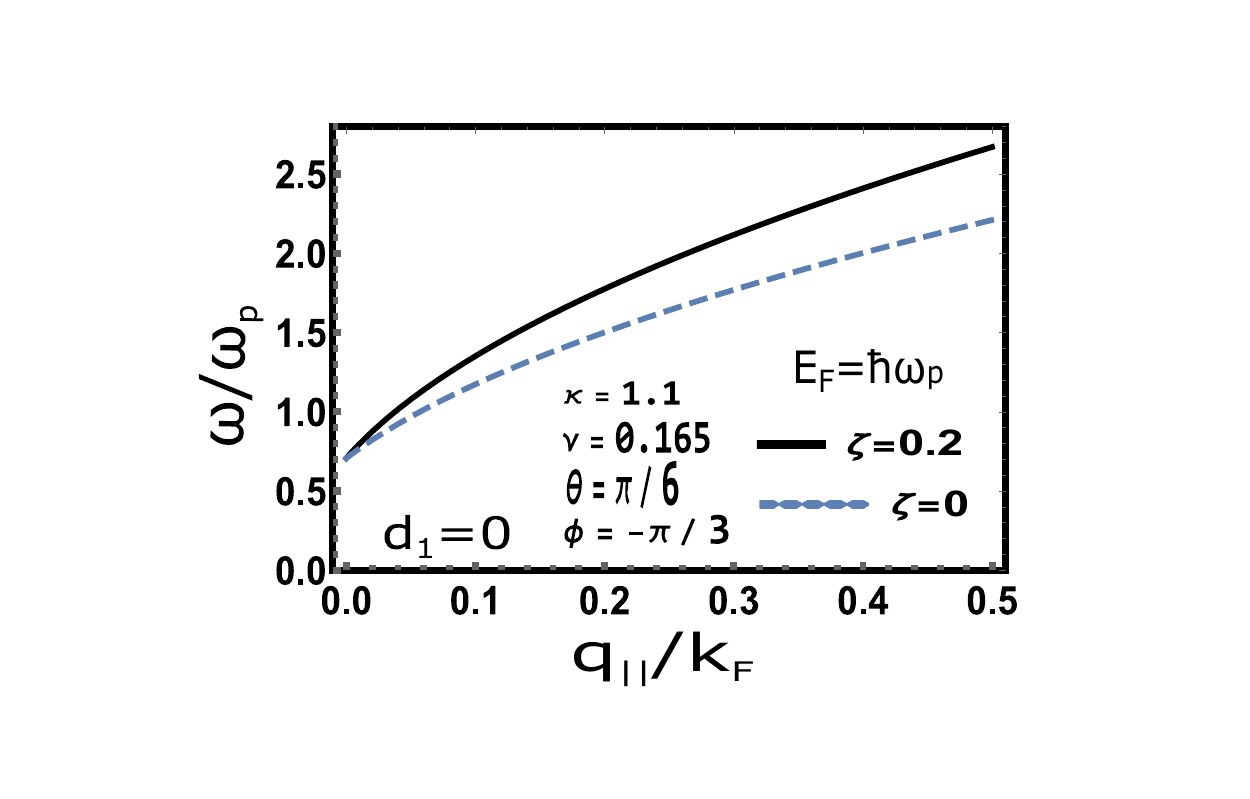}
\caption{(Color online) Plasmon mode dispersion for graphene sheet lying 
in contact with the conducting substrate. }
\label{FIG:2}
\end{figure}

When the plasmon-phonon interaction is turned off, the spectrum of plasmon branches is 
changed drastically. In Fig.\ \ref{FIG:Hornopho}, we present results for the plasmon 
mode dispersion for a structure consisting of a  graphene layer separated from the 
conducting substrate by a distance $d_1$. The space in between them is filled with a 
dielectric having background constant $\epsilon_b=2.4$, the known value for bulk 
graphite.\cite{Tando}   This separation gives rise to a linear low-frequency ``acoustic" mode 
similar to the one in Fig.\ \ref{FIG: plasphoint}(b) for which there is also a spacer-layer in the structure.  In Fig.\  \ref{FIG: plasphoint}(a), there is also a
plasmon branch  which is a hybrid with the surface plasmon with frequency
	$\omega_p/\sqrt{2}$ in accordance with Eq.\ (\ref{number1}).   A previous paper\cite{Horing} for 
	unstrained graphene interacting with a conducting substrate has also demonstrated 
	the existence of two modes similar to those  appearing in Fig.\  \ref{FIG:Hornopho}(a). 
	However, our main goal in presenting  Figs. \ref{FIG: plasphoint} 	and \ref{FIG:Hornopho} 
	is to show the influence of strain as well as plasmon-phonon interaction for the described 
	structure we are investigating.
To present the matter in more detail, we have displayed the variation of the plasmon frequency 
with change in the direction of the applied strain in Fig. \ref{FIG:Hornopho}(b). The plots  
show that for chosen wave vector and a specified direction of the applied strain, we have two 
plasmon frequencies. The one with a lower frequency  corresponds to acoustic plasmon whereas the 
higher frequency branch corresponds to hybrid plasmon mode.  The plots  also  illustrate 
that the range of variation of both plasmon mode frequencies, keeping the magnitude and 
direction of the strain fixed and for chosen small $q_\parallel$.  					
  In  Fig. \ref{FIG:Hornopho}(b), the intersection of two  plasmon branches implies that for different 
	directions of applied strain we can have the same resonating frequency for a 
	same direction of the wave vector.
We also show in Fig. \ref{FIG:2} how the plasmon spectrum in Fig.\ \ref{FIG:Hornopho} 
gets affected when the separation between the 2D layer and the surface reduced to 
zero. In any case, we still keep the interaction between the 2D layer and plasmons in the 
substrate. The resulting spectrum consists of only one branch originating near the surface 
plasmon frequency, $\omega_p/\sqrt{2}$, as is well known.\cite{Boperson} The figure demonstrates the significant role in modification of the plasmon branch slope due to the application of strain although the linearity of the dispersion curve in the long wavelength limit is still preserved. We also observe only one plasmon branch which in comparison to Fig. \ref{FIG: plasphoint}(a) shows the disappearance of plasmon mode rooted from $\omega_p/2$ as an important effect of absence of plasmon phonon interaction.

\section{Plasmon excitations for a graphene-2DEG double layer}
\label{insert}

  In a  recent paper, Politano, et al. \cite{Politano} reported  some interesting
results for the plasmon excitations when graphene weakly interacts with a Cu(111) 
substrate.  Momentum-resolved electron-energy-loss spectroscopy used in their 
experiments revealed multiple ``{\em acoustic\/}" surface plasmons. These authors 
accounted for this occurrence of low-frequency plasma modes as 
arising from both the graphene overlayer and the Cu(111) substrate 
If we follow the paper of Ahn, et al. \cite{PhysicaB} and treat the 
Cu(111) substrate as a 2DEG, this means that we may adopt our model as 
follows. There is a graphene overlayer with vacuum on one side   and a 
semi-infinite dielectric with constant $\epsilon_b$ on the other.  We have 
embedded in this dielectric a 2DEG at a distance $d_1$ from the graphene
layer. A straightforward calculation renders the surface response function for 
this arrangement as

\begin{equation}
g(q_{\parallel},\omega)=\frac{\left[(\text{$\epsilon_b$}-1)-\frac{\text{$\chi_1$}}{q_\parallel \text{$\epsilon_0$}}\right] \left[2 \text{$\epsilon_b$}-\frac{\text{$\chi_2$}}{q_\parallel \text{$\epsilon_0$}}\right]- e^{-2q_\parallel \text{$d_1$} } \left[\frac{\text{$\chi_1 $}}{q_\parallel\text{$\epsilon_0$}}+(\text{$\epsilon_b$}+1)\right]\frac{\chi_2}{q_\parallel\text{$\epsilon_0$}}}
{\left[(\text{$\epsilon_b$}+1)-\frac{\text{$\chi_1$}}{q_\parallel\text{$\epsilon_0$}}\right] \left[2 \text{$\epsilon_b$}-\frac{\text{$\chi_2$}}{q_\parallel\text{$\epsilon_0$}}\right]- e^{-2 q_\parallel \text{$d_1$} } \left[\frac{\text{$\chi_1$}}{q_\parallel\text{$\epsilon_0$}}+(\text{$\epsilon_b$}-1)\right] \frac{\chi_2}{q_\parallel \text{$\epsilon_0$}}}\ ,
\label{newestG}
\end{equation}
Where   $ \chi_j(q_\parallel,\omega)/(q_\parallel \epsilon_0)\approx C_j q_\parallel / \omega^2$,  $C_j$  ($j=1,2$) is constant in the long wavelength limit. AS a result, one can
show that the poles of the surface response function in Eq.\ (\ref{newestG}) 
correspond to the plasmon frequencies

\begin{equation}
\omega_\pm(q_\parallel)= \frac{1}{2}\left\{\frac{1}{\epsilon_b(\epsilon_b+1)}e^{-2 q_\parallel d_1 }(C_2q_\parallel (\epsilon_b-1)+e^{2 q_\parallel d_1 }
q_\parallel (C_2+2 C_1 \epsilon_b+C_2 \epsilon_b)\pm R(q_\parallel))\right\}^{1/2}\ ,
\label{plusminus}
\end{equation}
where 

\begin{equation}
R(q_\parallel)=   q_\parallel \bigg [-8 C_1 C_2 e^{2q_\parallel  d_1   }(e^{2q_\parallel d_1 }-1)\epsilon_b(\epsilon_b+1)+\bigg\{C_2(\epsilon_b-1)+e^{2q_\parallel d_1 }\big(C_2+2 C_1 \epsilon_b+C_2 \epsilon_b\big)\bigg\}^2\bigg]^{1/2}\ .
\end{equation}
The dispersion relation in Eq.\ (\ref{plusminus})  is interesting and needs to be 
analyzed in some detail.  If $C_1\neq C_2$, as is most likely the case 
for a graphene-2DEG double layer, then {\em both\/} modes  have 
a $\sqrt{q_\parallel}$ behavior at long wavelengths. However, if
$C_1=C_2=C$ and both layers are embedded in  a medium with 
 uniform  background dielectric constant, as was the case in Ref.\
[\onlinecite{DasMad}], then the frequencies are given by

\begin{equation}
\omega_{\pm}^2(q_\parallel) \approx \frac{Cq_\parallel}{2\epsilon_b}\left(1 \pm 
e^{-q_\parallel d_1} \right)
\end{equation}
so that {\em one\/} mode has  a
$\sqrt{q_\parallel}$ dependence while the other is linear in $q_\parallel$.
All of this is incumbent on the appearance  of the 2D Fourier transform of the 
Coulomb interaction as $2\pi e^2/(4\pi \epsilon_0  q_\parallel)$ which appears 
naturally in the procedure used for calculating the surface response function.  
In the paper of Ahn, et al. \cite{PhysicaB}, a screening parameter is introduced
into the 2D Fourier transform of the Coulomb potential, which has no place in 
our calculations. In summary, the fundamental
differences in the plasmon dispersion relations stemming from  Eq.\ (\ref{plusminus})
arise from the nonlocal screening by the background as well as the  hybridization of 
the underlying 2D modes.

\section{Numerical Results and Discussion}
\label{sec5}
	
\subsection{Plasma Excitations for gapless graphene}

\begin{figure}
\centering
\includegraphics[width=0.75\textwidth]{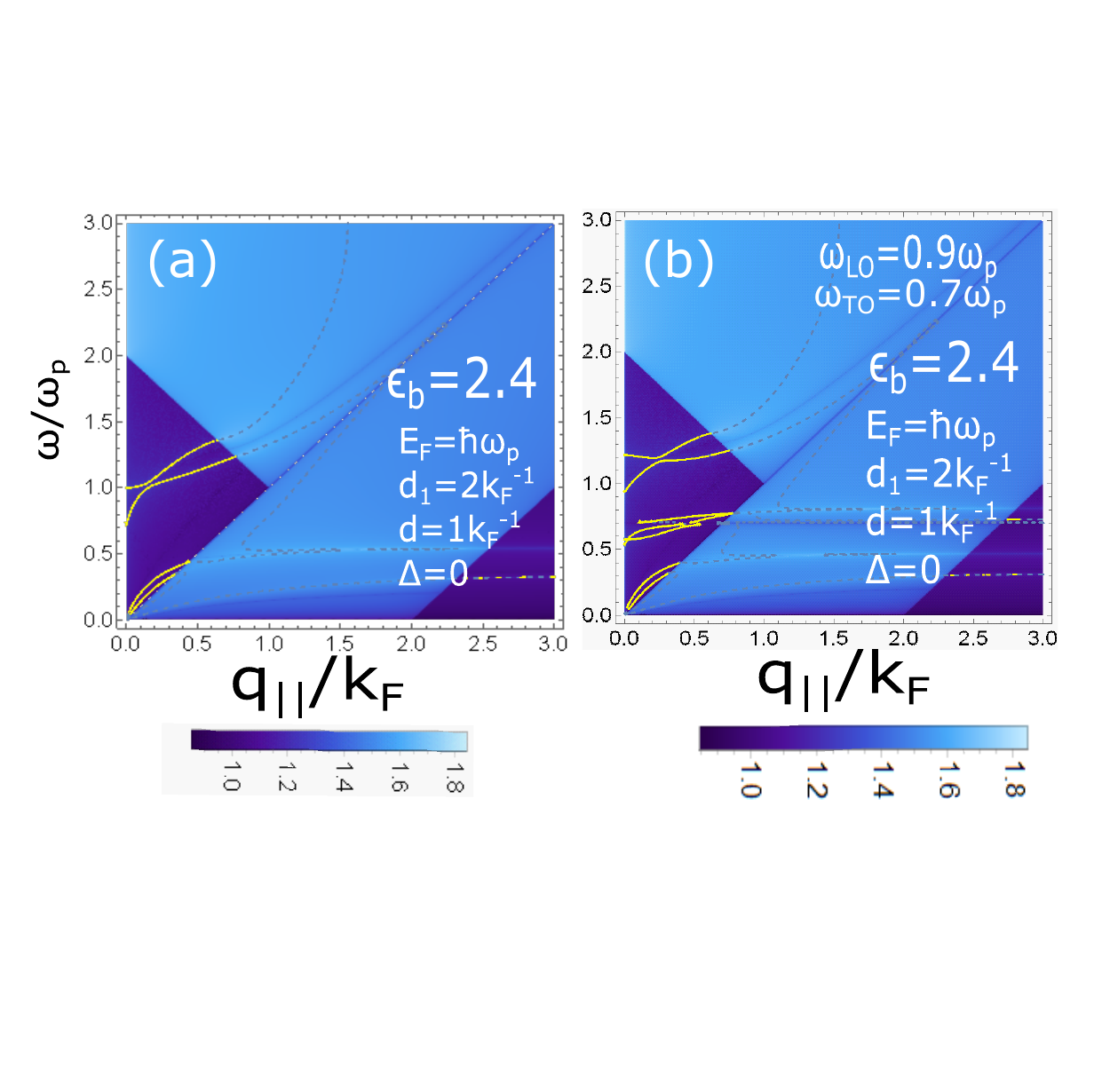}
\caption{(Color online)   Plasmon excitation spectra for gapless graphene. In
(a), the plasmon-phonon coupling is neglected. In (b), the plasmon-phonon
interaction is included.}
\label{FIG:4}
\end{figure}
 By making use of the expression for the surface response function in Eq.\  (\ref{gfn}),
we have carried out numerical calculations to obtain the plasmon dispersion relation for the  hybrid structure shown in Fig.\ \ref{FIG:1}.  The plasmon modes can be clearly seen in Fig. \ref{FIG:4} where our results are presented as density plots. These results illustrate  the plasmon mode for a pair of gapless graphene layers with one of them serving as a  protective layer on top and the other embedded within a medium of dielectric constant $\epsilon_b=2.4$. We have chosen an encapsulating
dielectric material with dielectric function $\epsilon_1(\omega)$. The plot in the left panel of Fig. \ref{FIG:4} shows the plasmon spectrum in the absence of phonon effects.
In this case, we observe four plasmon modes in Fig. \ref{FIG:4}(a), two of which originate from the origin and are due to the 2D plasmon modes ($\omega\sim q_\parallel^{1/2}$)  of free-standing graphene. The remaining two have frequencies  which are shifted by a depolarization from the bulk plasma frequency $\omega_p$ of the pair of encapsulating dielectric materials. However, Fig. \ref{FIG:4}(b) shows the plasmon excitations due to plasmon phonon interaction.  In Fig. \ref{FIG:4}(b), 
we observe two additional plasmon branches along with the four plasmon modes in Fig. \ref{FIG:4}(a). These two new plasmon modes are the result of longitudinal and
transverse optical phonon modes which couple with the graphene plasmon mode. In both Figs. \ref{FIG:4}(a) and (b), the plasmon modes get Landau damped as soon as they enter  the single-particle excitation region(light blue). Sharp boundaries  could be seen defining these regions in the figure.

\subsection{Plasma Excitations for gapped graphene}

\begin{figure}
\centering
\includegraphics[width=0.75\textwidth]{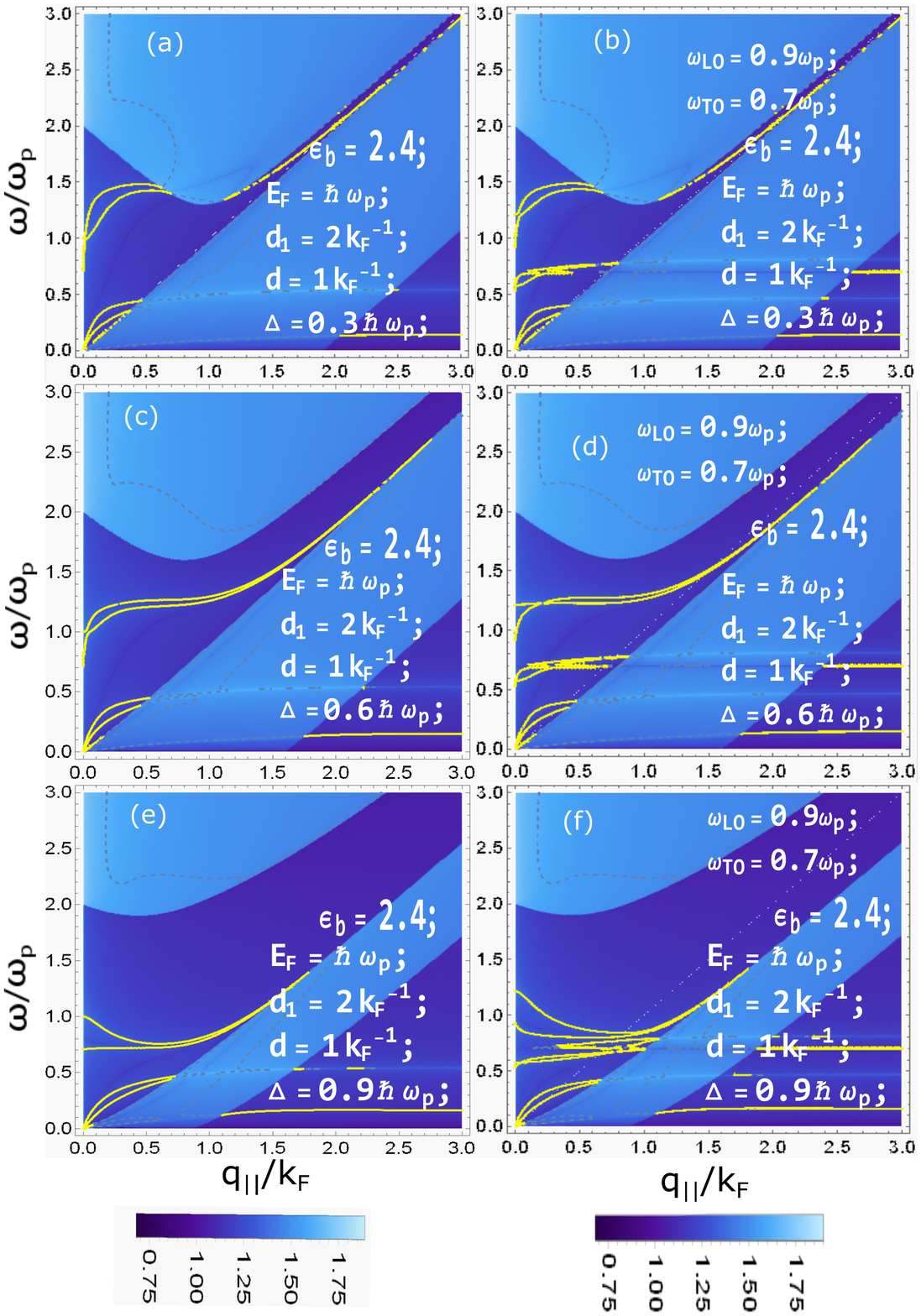}
\caption{(Color online)   Plasmon excitation spectra for gapped graphene. In the left panel Fig
(a), (c) and (e), the plasmon-phonon coupling is neglected. In right panel, Fig. (b), (d) and (f), the plasmon-phonon
interaction is included.}
\label{FIG:5}
\end{figure}

In Fig. \ref{FIG:5}, we present our results which show the influence on the plasmon mode dispersion arising from  lattice vibrations in the substrate  for the structure shown in Fig. \ref{FIG:1} when the used graphene layers have an energy band gap described by the parameter $\Delta=0.3,\ 0.6,\ 0.9\hbar\omega_p$. The figures in the left panel show two pairs of plasmon modes: one pair arising  from the origin and the other pair near the bulk plasmon frequency whereas the figures on the right panel show three pairs of plasmon modes. This additional pair which lies in between the other upper and lower plasmon modes is a direct result of the plasmon phonon coupling. For comparison with Fig. \ref{FIG:4}, we chose the same values of parameters for the transverse and longitudinal optical phonon frequencies $\omega_{LO}$ and $\omega_{TO}$, the  static background dielectric constant $\epsilon_b$, the doping level as well as the thickness of the encapsulating materials. The density plots in both left and right panels show that due to the introduction of the band gap, the particle-hole excitation region splits into two parts creating a region where plasmon mode can be excited as damping- free self-sustained charge density oscillations. This region widens with the increase of the  band gap leading to expanded regions for the charge density to oscillate without Landau damping. Due to increasing band gap, the members from each pair of plasmon mode group begin to merge at larger wave vector corresponding to short-range coupling. 

\subsection{Plasma Excitations for strained graphene}

\begin{figure}
\centering
\includegraphics[width=0.65\textwidth]{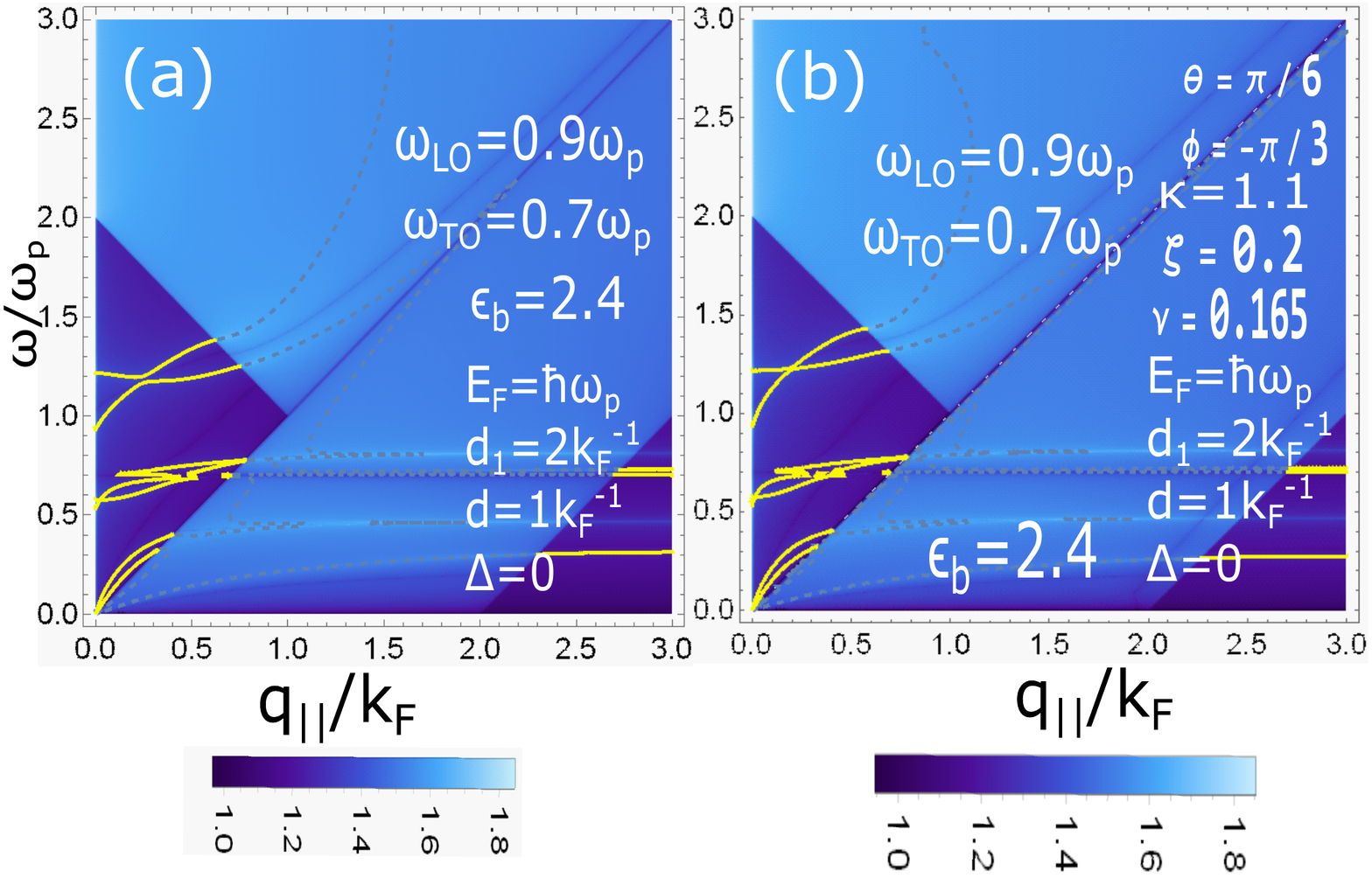}
\caption{(Color online)   Plasmon excitation spectra for graphene 
showing the effect of strain and phonon vibration on the plasmon dispersion with one 
layer of graphene as a protective layer and the other layer of graphene sheet
encapsulated in between dielectric materials.  Panel (a) shows the effect due to
coupling of plasmons with phonons when gapless graphene is considered.
Panel (b) shows the coupling of plasmon with phonon when the gapless graphene is under strain.}
\label{FIG:6}
\end{figure}

We have carried out additional calculations to examine the effect due to strain on the plasmon mode dispersion of graphene layers.  In Fig.\  \ref{FIG:6}, we have presented our numerical results to illustrate the strain effect in the presence of plasmon phonon coupling for the given heterostructure  shown in Fig. \ref{FIG:1}. The left panel of Fig.\  \ref{FIG:6} shows the plasmon mode in the absence of strain whereas the right panel displays cases in the presence of strain. No great difference can be observed between these two panels because the application of a small strain does not affect the energy band structure noticeably.  A small distinction between them is on the upper most pair of plasmon modes. Here, we see the two modes are not in contact with each other under no strain but they come closer under a finite strain. In a recent paper,\cite{Merano}  a generalization of the early surface
plasmon theory [see Ref.\ \onlinecite{Raether}],  was presented by including a surface current\cite{Huang}  flowing within either a graphene or a Boron-Nitride monolayer on the surface of a bulk dielectric. Although the retarded interaction between the incident light and electrons in a monolayer was employed for calculating surface confinement of the TE mode of light and its propagation loss, the important nonlocal dynamics 
involved in optical response of electrons \cite{Iurov}  was  neglected. Under strain, we 
anticipate that this mode will be affected. 

\subsection{Contributions to Energy Loss}

In Sec.\ \ref{sec2}, we demonstrated that the power loss for a beam of charged
particles moving with velocity ${\bf v}$ at a distance $z_0$ from a surface
may be expressed  in terms of $\mbox{Im}\ g\left({\bf q_\parallel},\omega
={\bf q}_\parallel  \cdot  {\bf v}\right)$ as given in Eq.\ (\ref{FINAL}).  Then,
subsequently,  in Eq.\ (\ref{gfn}), we expressed the surface response function
in fractional form as
$g({\bf q}_{\parallel},\omega)= {\cal N}({\bf q}_{\parallel},\omega)
/{\cal D}({\bf q}_{\parallel},\omega)$. We may separate ${\cal N}={\cal N}_R+i{\cal N}_I$
and ${\cal D}={\cal D}_R+i{\cal D}_I$  into their real and imaginary parts so that

\begin{equation}
\mbox{Im}\  g\left({\bf q_\parallel},\omega \right)
={\cal N}_I \left(\frac{{\cal D}_R}{{\cal D}_R^2+{\cal D}_I^2}  \right)
-{\cal N}_R \left(\frac{{\cal D}_I}{{\cal D}_R^2+{\cal D}_I^2}  \right) \  .
\label{FORM}
\end{equation}
 Given the form in Eq. \ (\ref{FORM}), there is a contribution to the integrand
in Eq.\ (\ref{FINAL})  whenever we have either (a)\
${\cal D}_I({\bf q}_\parallel,\omega={\bf q}_\parallel\cdot{\bf v})\neq 0$ or
(b) both ${\cal D}_I({\bf q}_\parallel,\omega={\bf q}_\parallel\cdot{\bf v}) $ and
  ${\cal D}_R({\bf q}_\parallel,\omega={\bf q}_\parallel\cdot{\bf v}) $ are
simultaneously equal to zero.  When case (a)  holds, we have Landau damping and the
particle-hole region contributes to the energy loss.  In case (b),  the dispersion equation
for plasmon excitations is satisfied in the hybrid structure and the plasmon modes
contribute.  In this case, we use the Dirac identity so that

\begin{equation}
\mbox{Im}\  g\left({\bf q_\parallel},\Omega_p \right)=
  \pi {\cal N}_I
\frac{\delta(\Omega_p-{\bf q}_\parallel \cdot{\bf  v})}{\left|
\partial {\cal D}_I/\partial\omega  \right|}
-\pi {\cal N}_R
\frac{\delta(\Omega_p-{\bf q}_\parallel \cdot{\bf  v})}{\left|
\partial {\cal D}_R/\partial\omega  \right|}
\end{equation}
where the derivative here is to be evaluated at the plasmon frequency $\Omega_p$. In the case, when 
a graphene layer is free-standing and embedded in a dielectric medium, the power loss is 
simplified for a high-speed charged particle and given by

\begin{equation}
\frac{dW}{dt}=\frac{Z^\ast e^2}{8\epsilon_0} \int_0^\infty  dq_\parallel  \int_0^{\pi/2}  d\alpha \  e^{-2q_\parallel z_0}
\frac{ \sqrt{K B(\theta,\phi)}\left| \sqrt{KB(\theta,\phi)q_\parallel} +q_\parallel v\cos\alpha \right|}{\sqrt{q_\parallel} v\cos\alpha}
\delta\left( 1-\frac{KB(\theta,\phi)}{q_\parallel v^2\cos^2\alpha}  \right)\ .
\end{equation}

 \begin{figure}
\centering
\includegraphics[width=.75\textwidth]{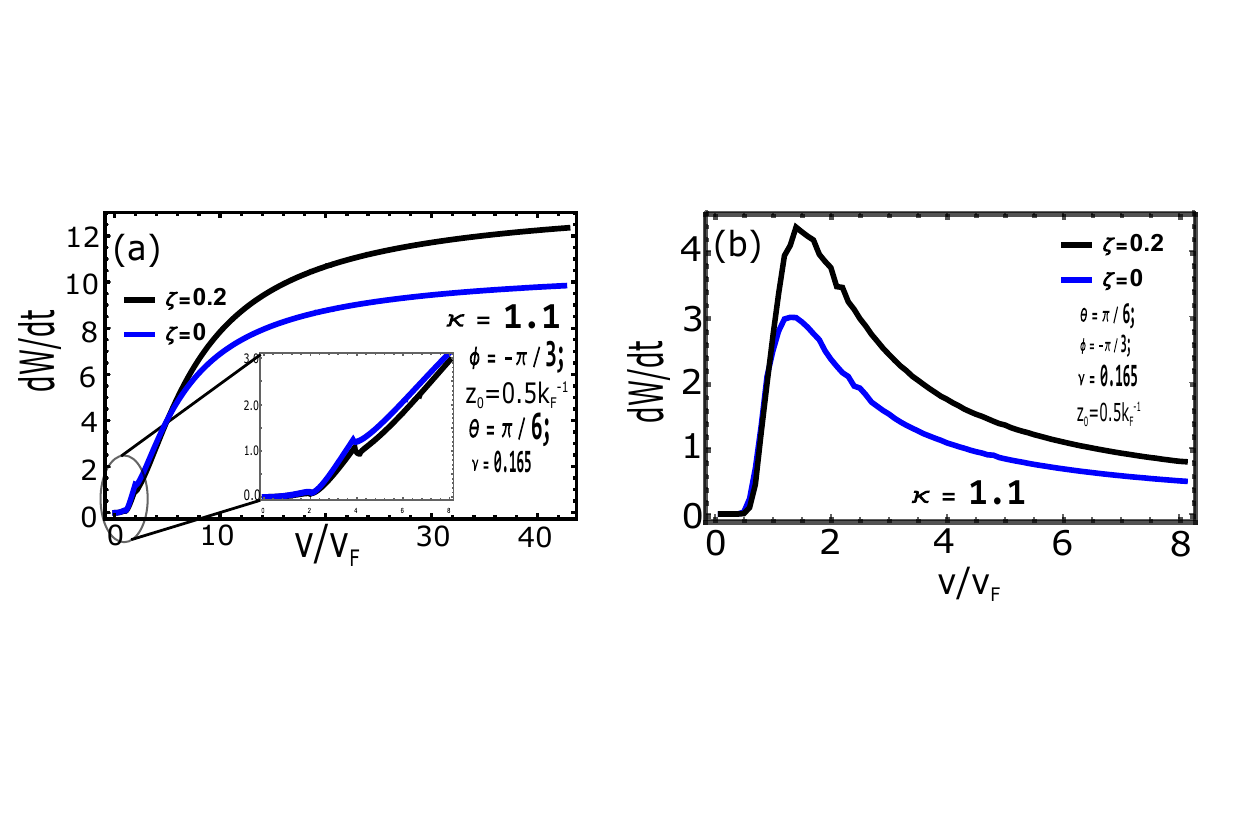}
\caption{(Color online) The plots  show the energy loss rate in units of $Z^\ast e^2k_F^2v_F/(8 \epsilon_0)$
due to (a)  paricle hole mode and (b)  the plasmon excitations
 for a freely suspended strained and unstrained graphene.}
\label{FIG:a}
\end{figure}

 \medskip
\par

Making use of Eq.\. (\ref{FINAL}) and the surface response function in  Eq.\  (\ref{gfn}) for free-standing  graphene, we have numerically calculated the contributions to the rate of loss of energy for a charged particle, moving parallel over the graphene sheet, due separately to single-particle excitations and the plasmon modes.  Our results shown in Fig.\ \ref{FIG:a} simply present the variation of the rate of loss of energy as a function of the impinging particle velocity for a chosen height $z_0=0.5k_F^{-1}$. Comparison of plots for strained and unstrained graphene shows that the results are qualitatively similar over the exhibited velocity range. However, a distinct difference is observed in their magnitudes. At low  velocities of a charged particle,   the energy loss rates for both strained and unstrained graphene are almost equal.  But, at high velocities of an incoming  charged particle, the energy loss rate due to particle-holes and plasmon modes is enhanced for strained than for unstrained graphene. The energy loss rate due to particle-hole modes is increased first and eventually levels off as the value of the charged particle velocity is raised. On the other hand,     the energy loss rate due to plasmon excitations for either strained or unstrained graphene remains negligible at small velocity and beyond a critical  value it increases rapidly to a maximum after which it starts decreasing continuously as the particle velocity becomes larger and larger.   Overall, the energy loss rate for strained graphene is greater than for unstrained graphene.

\subsection{Screened Impurity potential}

Starting with Eq.\ (\ref{eq:8}), we obtain the static screening of the potential on the surface at $z=0$ 
due to an impurity with charge $Z_0^\ast e$ located at distance $z_0$ above the surface of the hybrid structure shown in Fig.\ \ref{FIG:1}.
We have

\begin{equation}
\phi ({\bf r}_\parallel,\omega=0) =  \frac{Z_0^\ast e}{2\pi \epsilon_0} \int_0^\infty  dq_\parallel \int_{0}^{2\pi} d\theta\
e^{iq_\parallel r_\parallel \cos\theta}
\left[ 1-g({\bf q_\parallel},\omega=0)  \right] e^{-q_\parallel z_0}\ .
\label{eq:8scr}
\end{equation}

\begin{figure}
\centering
\includegraphics[width=.75\textwidth]{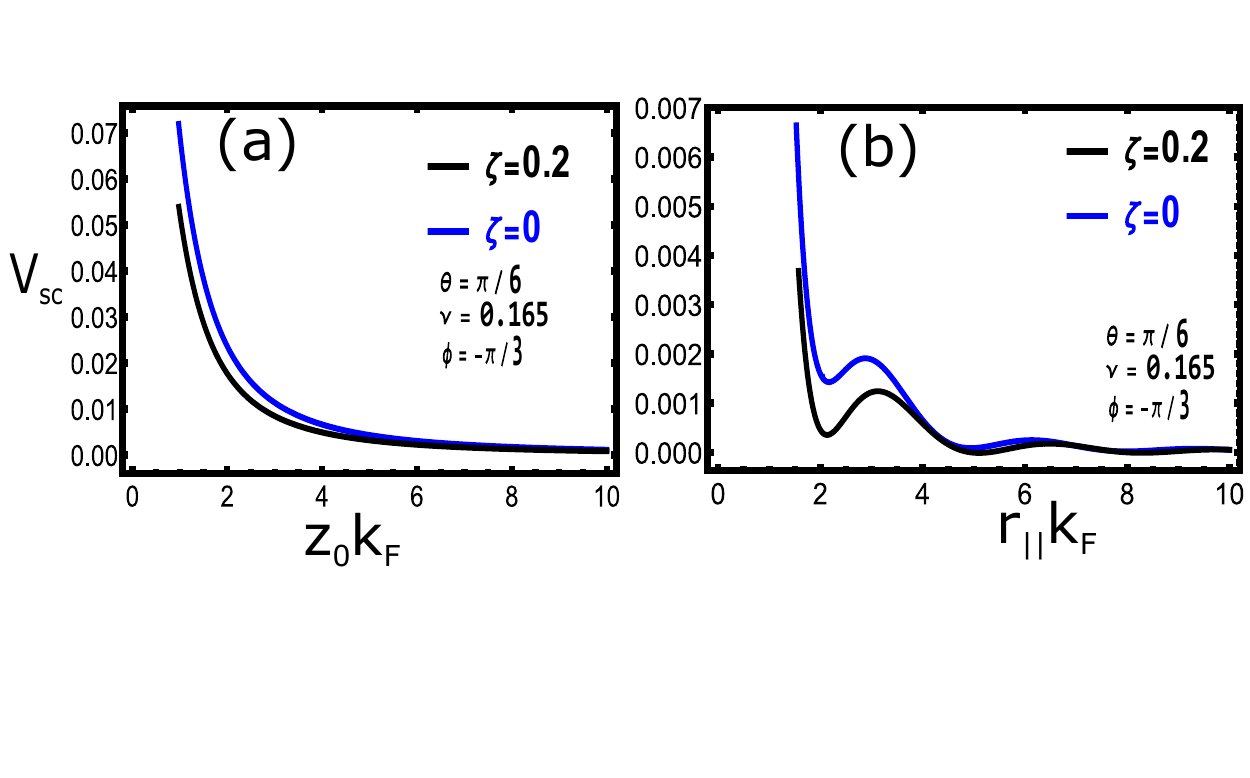}
\caption{(Color online) The screened impurity potential in unit of $Z_0^\ast e k_F/(2\pi\epsilon_0)$ is plotted as a function of (a) height $z_0$ and (b) $r_{\parallel}$ respectively for strained and unstrained graphene monolayers for the chosen parameters in the figure. }
\label{FIG:13}
\end{figure}
\newpage

By employing Eq.\ (\ref{eq:8scr}), we have computed the screened impurity potential $V_{sc}$ for both strained and unstrained monolayer graphene. In Fig.\ \ref{FIG:13}(a), the screened potential decays exponentially with increasing height $z_0$.  This behavior applies for both strained and unstrained graphene.  However, we note  that there is a significant variation in the screened potential   when the charge is put closer to the graphene sheet. In Fig.\ \ref{FIG:13}(b), we have calculated the screened potential as a function of the in-plane variable   $r_{\parallel}$ in units of the inverse Fermi wave number for both strained and unstrained graphene. The plot shows the occurrence of Friedel oscillations with the potential being shifted upward when strain is applied. We also notice that there is no significant change in the screened potential for strained and unstrained graphene as long as the value of $r_{\parallel}$ is large.

\medskip
\par

\section{  Concluding Remarks}
\label{sec6}

We have determined an expression for the rate of loss  of energy for a beam of charged particles traveling parallel to the surface of a hybrid structure explicitly in terms of its surface response function.  The formalism covers the case when the dependence of the response function on the in-plane wave vector is anisotropic.  Specifically, we apply our formalism to investigate uniformly strained graphene both analytically and numerically.  
We report on the low-energy plasma excitations using an effective Dirac Hamiltonian which reveals the absence of graphene trigonal symmetry at the lowest order for weak strain.  In particular, we investigate and report results for the effect due to the  deformation of the Dirac cone,  the  band gap, doping level, thickness of the substrate, screening due to dielectric material and the effect of plasmon-phonon coupling on the plasmon modes,   energy loss and static shielding of an  impurity located either just outside  the surface of the hybrid structure
or embedded inside it. The versatility of our calculated results is that it governs an extended range of applications for investigating impurity shielding, power loss of impinging charged particles as well as the charge density oscillations for a hybrid structure such as the one depicted in Fig.\ \ref{FIG:1}. Strained graphene may be successfully  substituted by alternative 2D materials having planar or buckled structures with lattice asymmetry. With the use of our procedure,  a wide variety of stacking arrangements may also be adapted.

\acknowledgments
D.H. would like to thank the support from the Air Force
Office of Scientific Research (AFOSR). DH is also supported by the DoD Lab-University Collaborative Initiative (LUCI)Program.

\end{document}